\documentclass[final, runningheads]{llncs}
\usepackage[T1]{fontenc}
\usepackage{graphicx}
\usepackage[utf8]{inputenc} \usepackage[T1]{fontenc}    \PassOptionsToPackage{hyphens}{url}
\usepackage{hyperref}       \usepackage{url}            \usepackage{hyperref}
\usepackage{booktabs}       \usepackage{amsfonts}       \usepackage{nicefrac}       \usepackage{microtype}      \usepackage{lipsum}
\usepackage{fancyhdr}       \usepackage{graphicx,subcaption}
\usepackage{booktabs}
\usepackage{makecell}
\usepackage{enumerate}
\usepackage{enumitem}
\setlist[itemize]{noitemsep, nolistsep}
\usepackage{floatrow}
\usepackage[toc]{appendix}

\newfloatcommand{capbtabbox}{table}[][\FBwidth]

\usepackage{caption}
\graphicspath{{media/}}     \usepackage{amsmath}

\usepackage{algorithm}\usepackage{algpseudocode}

\usepackage{amsmath}
\newcommand{\itadata}{\footnotesize \textsl{ITADATA2024: The 3$^{\text{rd}}$ Italian Conference on Big Data and Data Science}}
\usepackage{fancyhdr}
\fancypagestyle{empty}{\fancyhf{}\fancyhead[C]{\itadata}
  \fancyhead[R]{\includegraphics[width=.05\textwidth]{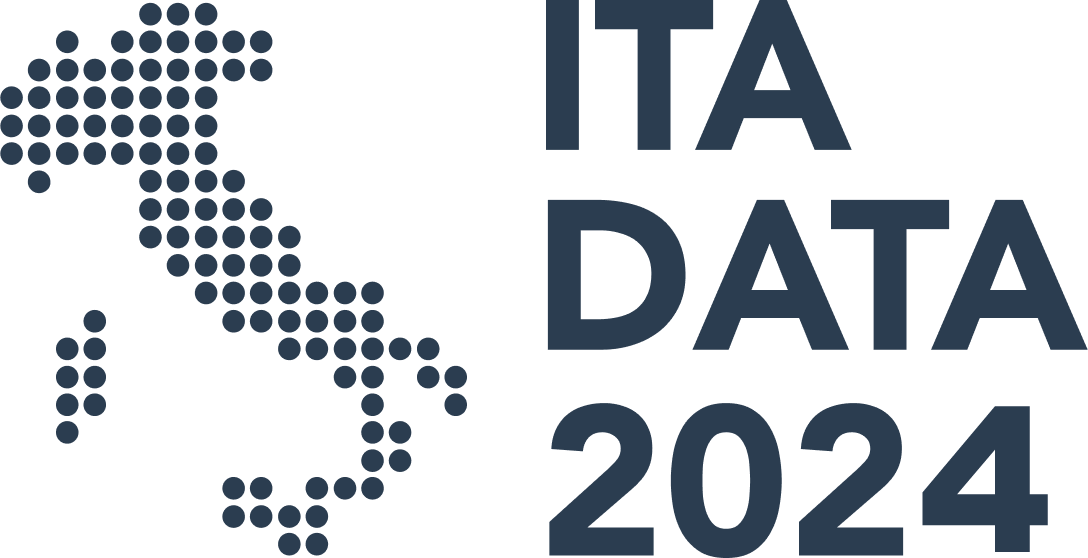}}}

\makeatletter
\begin{document}
\title{WeirdFlows: Anomaly Detection in Financial Transaction Flows}
\author{Arthur Capozzi\inst{1}\and
Salvatore Vilella\inst{1,2}\and
Dario Moncalvo\inst{3} \and Marco Fornasiero\inst{3} \and Valeria Ricci\inst{3} \and Silvia Ronchiadin\inst{3}
\and Giancarlo Ruffo\inst{2}}
\authorrunning{A. Capozzi et al.}
\institute{Dipartimento di Informatica,
  Universit\`a degli Studi di Torino,
  Torino, Italy \\\email{{name}.{surname}@unito.it}  \and
DISIT,
  Universit\`a degli Studi del Piemonte Orientale ``A. Avogadro'',
 Alessandria, Italy \\
\email{{name}.{surname}@uniupo.it}\\
\and Anti Financial Crime Digital Hub,
  Turin, Italy \\
  \email{{name}.{surname}@intesasanpaolo.com} }
\maketitle              \begin{abstract}
In recent years, the digitization and automation of anti-financial crime (AFC) investigative processes have faced significant challenges, particularly the need for interpretability of AI model results and the lack of labeled data for training. Network analysis has emerged as a valuable approach in this context.

In this paper we present \textit{WeirdFlows}, a top-down search pipeline for detecting potentially fraudulent transactions and non-compliant agents. In a transaction network, fraud attempts are often based on complex transaction patterns that change over time to avoid detection. The \textit{WeirdFlows} pipeline requires neither an a priori set of patterns nor a training set. In addition, by providing elements to explain the anomalies found, it facilitates and supports the work of an AFC analyst.

We evaluate \textit{WeirdFlows} on a dataset from Intesa Sanpaolo (ISP) bank, comprising 80 million cross-country transactions over 15 months, benchmarking our implementation of the algorithm. The results, corroborated by ISP AFC experts, highlight its effectiveness in identifying suspicious transactions and actors, particularly in the context of the economic sanctions imposed in EU after February 2022. This demonstrates \textit{WeirdFlows}' capability to handle large datasets, detect complex transaction patterns, and provide the necessary interpretability for formal AFC investigations.

\keywords{Network Analysis \and 
Anti Financial Crime \and Anti Money Laundering \and Temporal Networks \and Graph Search Algorithm}
\end{abstract}
\section{Introduction and Related Work}

Financial markets are complex, adaptive systems involving diverse actors such as hedge funds, individual investors, and banks. Complex networks effectively model these systems, focusing on stabilization and destabilization dynamics~\cite{Lillo_2008}, stock price correlations~\cite{Vandewalle}, shareholder networks~\cite{Caldarelli}, and the resilience of financial networks~\cite{Peron}. International trade networks based on linguistic, cultural, or religious ties have also been examined~\cite{Rauch01}, along with applications to financial crises, including the 2008 global financial crisis~\cite{FAGGINI2019105}. Beyond stock markets, network theory models competitive dynamics on the World Wide Web~\cite{LOPEZ2003754} and the cryptocurrency market, analyzing dominant cryptocurrencies~\cite{Papadimitriou}, transaction structures~\cite{Serena}, and Ethereum network attributes~\cite{LinDan}.

Network analysis is vital for anti-financial crime (AFC) and anti-money laundering (AML), detecting anomalies in transaction networks and automating fraud detection to enhance efficiency. Tools like VISFAN~\cite{VISFAN} use network metrics for identifying suspicious transactions. García et al.~\cite{García2021} applied network analysis in the Spanish Revenue Agency's Tax Control Study, using algorithms for rapid fraud detection and community detection techniques for economic landscape representation. Colladon et al.~\cite{FRONZETTICOLLADON201749} highlighted social network metrics for identifying money laundering through relational graphs of economic sectors, regions, transaction volumes, and ownership links. The CoDetect framework~\cite{HuangMu} integrates network and feature data for fraud detection.

AMLSim~\cite{Weber2018} generates synthetic bank transaction data with identifiable money laundering patterns. Preliminary results indicate that graph learning for AML is feasible even in large, sparse networks, with graph compression techniques like Ligra+ achieving significant compression~\cite{7149297}. Garcia-Bedoya et al.~\cite{Bedoya2020} advocate for AI and network analysis in AML, identifying three interaction types that conceal money laundering: path (money sent through intermediaries), cycle (money returns to its origin), and smurf (dividing a transaction into smaller ones). Liu et al.~\cite{Liu_Zhou_Zhu_Gu_He_2020} suggest transaction cycles can indicate fraud.

Machine learning and deep learning, particularly graph neural networks, are explored for fraud detection~\cite{Chen2018}. Unsupervised anomaly detection is often used due to a lack of annotated training data, with innovative approaches like zero-shot learning~\cite{Chen2018}. Pan~\cite{Pan2022} proposes a deep-set algorithm combining meta-learning and zero-shot learning for money laundering detection. Reviews highlight effective anomaly detection strategies in fraud detection~\cite{boltonhand2002,phuaetal2010}, using social network analysis to uncover organized criminal behavior~\cite{SUBELJ20111039}.

\subsection{WeirdFlows}

In this paper, we introduce \textit{WeirdFlows}, a top-down search pipeline based on network analysis, designed to aid Anti-Financial Crime (AFC) analysts in detecting illicit transactions and non-compliant agents. We evaluate \textit{WeirdFlows} on a dataset of 80 million cross-country bank transactions over 15 months, provided anonymously by Intesa Sanpaolo bank, compliant with legal privacy and security standards. This evaluation focuses on identifying attempts to bypass economic sanctions against Russia following the Ukraine invasion on February 24, 2022.

Financial fraud often involves complex, evolving transaction patterns to evade detection. \textit{WeirdFlows} identifies these by defining a \textit{transaction flow} as a set of payment lines from $x$ to $y$ through intermediaries. Unlike other methods, \textit{WeirdFlows} does not rely on predefined patterns, addressing the challenge of unlabeled data in AFC tools. Moreover, it provides interpretability, crucial for formal investigations by domain experts, unlike many black-box models~\cite{Chen2018}.

The application of \textit{WeirdFlows} involves two steps:
\begin{enumerate}
    \item \textbf{Determine Structural Granularity}: create a transaction network using financial transactions data. Each node can be any sort of entity, such as an \textit{ISO code}, \textit{BIC}, or \textit{IBAN}. ISO 3166 codes define country names, BIC (Bank Identifier Code) identifies banks in international payments, and IBAN (International Bank Account Number) identifies bank accounts across borders.
    \item \textbf{Compute paths on temporal networks}: Generate multiple weighted directed temporal networks by aggregating wire transfers over time (e.g., weekly, monthly). An algorithm computes all possible paths from a selected node for each temporal aggregation. Analysts explore these paths to identify trends, patterns, or anomalies and estimate the total money sent between nodes through intermediaries.
\end{enumerate}

Section \ref{sec:results} presents the application of \textit{WeirdFlows}, demonstrating:
\begin{enumerate}
    \item Its effectiveness in identifying anomalous transaction flows;
    \item Its scalability to millions of transactions;
    \item Its interpretability, facilitating formal investigations.
\end{enumerate}

We formalize transaction networks and transaction flows in the next section, describe the algorithm for identifying transaction flows, and propose a technique for weighting flows between nodes. Sections \ref{sec:results} covers the application experiments and performance benchmarks of \textit{WeirdFlows} at different granularities, while the last section discusses the limitation and the possible extensions of the work.

\section{Methodology}
\label{subsec:methodology}

\subsection{Weighing a Transaction Flow}
\label{subsec:flowtheory}
Transactions networks, usually represented as weighted directed temporal networks, are very useful to model financial transactions over time. On such a network $G_T =(V, L_T)$, we can define 
a path as an ordered finite collection of $n$ distinct edges connecting vertices $i$ and $j$ during interval $T$:

\begin{equation} \label{eq: path}
\begin{split}
    P^T_{i, j} = \{ & e(i, v_1), e(v_1, v_2), e(v_2, v_3), \ldots, e(v_{n-1}, j) \},\\
    & \text{with } v_x \in V \text{ and } e(v_x, v_{x+1}) \in L_T \\
    & \text{for all } x \in \{1, \ldots, n-1\}
\end{split}
\end{equation}

The weight of the path $P^T_{i, j}$ is then given by \begin{equation}
    \sum_{i=1}^{n-1}w(e(v_i, v_{i+1}))  
\end{equation}

and $W(P_i) = \{w(e(v_1, v_2)), ..., w(e(v_{n-1}, v_{n}))\}$ is the set containing the weights of the edges of the path $P_i$. Finally, $\text{Paths}^n_{i, j}(G_T)$ is the set of all possible paths of length $n$ from $i$ to $j$ in graph $G$ at interval $T$.

In real-world scenarios, to avoid detection of fraudulent activity, an agent may use intermediaries to transfer money. The number of intermediary groups or the length of each path connecting any two nodes $x$ to $y$ often cannot be known a priori.

A transaction through $n$ intermediaries in a transaction network forms a path of length $n+1$. The \textit{WeirdFlows} pipeline aims to identify the transaction flow from a given input node $x$ and verify the maximum possible amount of money sent from $x$ to other nodes within a maximum distance $n$.

Equation \ref{eq: flow_def} defines $Flow^n(x, y)$ as the set of all paths from $x$ to $y$ with a maximum length $n$:

\begin{equation}\label{eq: flow_def}
    Flow^n(x, y) = \{P_1, \ldots, P_m\} \text{, with } P_i \in \text{Paths}^n_{i, j}(G_T)
\end{equation}

In a transaction network, the maximum amount node $x$ could send to node $y$ through intermediaries is the minimum weight of the edges in the path between $x$ and $y$. For multiple intermediary groups, the transaction flow weight between $x$ and $y$ is the sum of the minimum weights of each path. The flow weight of maximum length $n$ from $x$ to $y$ is defined as:

\begin{equation}\label{eq:flow_weight}
    w(\text{Flow}^n(x, y)) = \sum_{i=1}^{m} \text{min}(W(P_i))\text{, with } P_i \in \text{Paths}^n_{i, j}(G_T)
\end{equation}

\begin{figure}[!htp]
  \centering
  
  \subfloat[Figure]{\includegraphics[width=0.46\linewidth]{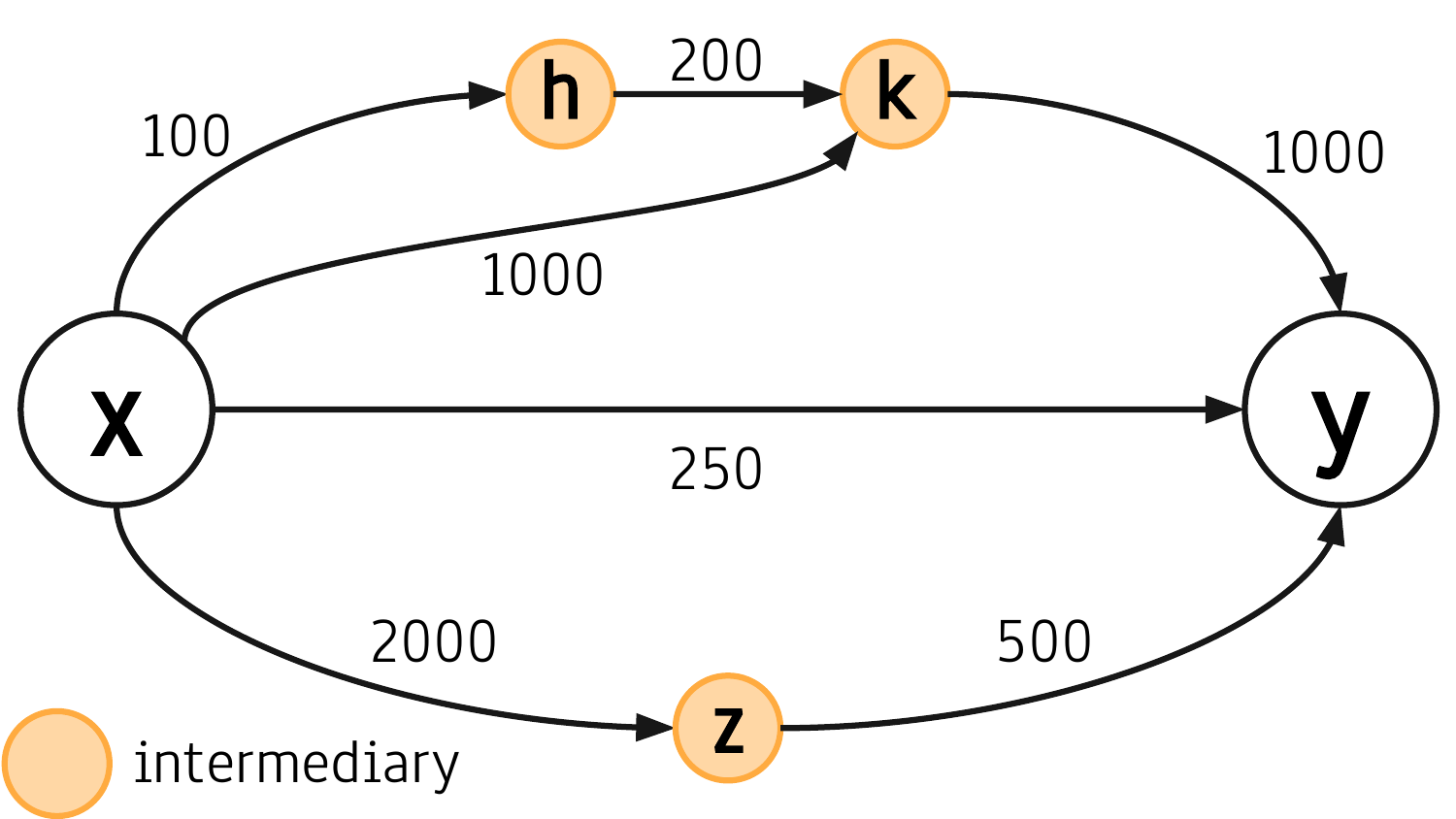}
    \label{subfig:net_example_2}
    }\hfill
  \subfloat[Table]{\begin{tabular}{l|l}
     \toprule
        $P_i$ & $min(W(P_i))$\\
         \midrule
         $\{e(x, h), e(h, k), e(k, y)\}$ & $100$\\
         $\{e(x, k), e(k, y)\}$ & $1000$\\
         $\{e(x, z), e(z, y)\}$ & $500$\\
         $\{e(x, y)\}$ & $250$\\
        \bottomrule
    \end{tabular}
    \label{subtbl:table_net_example_1}
  }
  \caption{In the transaction network in Figure \ref{subfig:net_example_2}, node $x$ has an edge of weight $250$ towards node $y$. The paths between nodes $x$ and $y$ through nodes $h$, $k$ and $z$ could be an attempt to hide a direct edge of higher weight. Table \ref{subtbl:table_net_example_1} lists all paths of maximum distance $3$ from $x$ to $y$ and their respective minimum weights.}
  \label{fig:table_net_example_2}
\end{figure}

Figure \ref{subfig:net_example_2} illustrates a transaction network. To study the hypothetical amount of money sent from $x$ to $y$, considering only the weight of edge $e_{x, y}$ is insufficient. Intermediaries like nodes $h$, $k$, and $z$ must be considered. Table \ref{subtbl:table_net_example_1} lists the paths and minimum weights for each path. The flow weight from $x$ to $y$ is the sum of the minimum weights for each path, totaling $1850$.

Note that the temporal aggregation of a network for the interval $T$ is an approximation. In fact, each transaction has a timestamp $t$ and therefore all edges involved in a valid path should satisfy the constraint $t_{e_i} < t_{e_{i+1}}$.

\subsection{Pseudocode}
\label{subsec:pseudocodice}

The pseudocode of the algorithm underlying \textit{WeirdFlows} is shown in \ref{algo:Flow_Algo}. It is a variant of the DFS graph search algorithm~\cite{dfs}, and it consists of a recursive search in a network $G$ of all paths of maximum length $n$ starting from a given input node $u$. The search explores the nodes in depth, starting from node $u$, and each search ends when the length of the path has reached the maximum length $n$. The Algorithm~\ref{algo:Flow_Algo} returns as output a list of lists, each representing a path and containing the ordered list of nodes traversed. It can be proven that the time complexity of the algorithm is less than or equal to $max(d_{out})^n (n+1)$, where $d_{out}$ is the out-degree of one of the nodes in the network. 

\begin{algorithm}
    \caption{WeirdFlows}\label{algo:Flow_Algo}
    \begin{algorithmic}[1]
        \Procedure{FindPaths}{$G, u, n, n\_recursive$}
            \State \texttt{paths}$\gets$ $[$ $]$
            \If{\texttt{n\_recursive} $ == 0$}
                \State \textbf{return} \texttt{[[u]]}
            \EndIf
            \If{\texttt{n\_recursive} $ < $ \texttt{n}}
                \State
                \texttt{paths.append([u])}
            \EndIf
            \For{\texttt{neighbor in G.neighbors(u)}}\Comment{loop through all neighbours of \texttt{u}}
                \For{\texttt{path in FindPaths(G, neighbor, n, n\_recursive - 1)}}
                    \If{\texttt{u not in path}}
                        \State \texttt{path.insert(0, u)} \Comment{append node \texttt{u} on the head of path list}
                        \State \texttt{paths.append(path)}
                    \EndIf
                \EndFor
            \EndFor
            \State \textbf{return} \texttt{paths}
        \EndProcedure
    \end{algorithmic}
\end{algorithm}

\subsection{Time Series of Flows}
\label{subsec:timeseriesflows}
Anomaly detection typically focuses on statistically significant changes in the weight of existing edges rather than identifying new edges. Detecting anomalies between two nodes at the path level involves discovering significant changes in weight, not new paths. In large financial transaction networks, connectivity is influenced by periodic payments forming repetitive patterns, such as salary payments or transfers to service providers. By analyzing a node's history, one can model transaction patterns and identify anomalies.

Applying \textit{WeirdFlows} in an AFC investigation involves examining the time series of $w(Flow)$. A flow comprises multiple paths, so a single edge-level anomaly may not significantly impact the overall flow weight.

Figure~\ref{fig:fig3} shows the time series of a transaction flow's weight between two nodes, with the flow weight in orange, the weighted moving average (WMA) in blue, and the exponentially weighted moving average (EWMA) in green. Anomaly detection can involve checking if the percentage difference between the expected value (WMA or EWMA) and the actual flow weight exceeds a threshold. In Figure~\ref{fig:fig3}, this difference is indicated by vertical dotted lines, green if positive and red if negative. Furthermore, an anomaly could be defined as exceeding a threshold percentage difference between the expected value from a time series forecasting model, like an Autoregressive Integrated Moving Average (ARIMA) model~\cite{box2015time}, and the actual flow weight; as well as considering seasonalities through (SARIMA) models~\cite{SARIMA}, though with high computational costs.

\subsection{WeirdFlows Pipeline}

We formalised the concepts described above in the \textit{WeirdFlows} pipeline, i.e., a sequence of steps that allow us to analyse all the possible flows of money between entities in a financial graph, that can be built at different levels of spatial and temporal aggregation. While this methodology can easily be generalised to any suitable graph, we will refer to financial graphs as this pipeline was originally designed to support an AFC investigation and we will illustrate its application specifically to financial graphs in this Section~\ref{sec:results}.

The \textit{WeirdFlows} pipeline can be described in the following steps:

\begin{enumerate}[noitemsep,topsep=0.3pt]
\itemsep0.3em
    \item Choose a structural and  temporal aggregation $T$ of the data in order to build the graph;
    \item For each temporal aggregation, create a weighted temporal transaction network $G_T$;
    \item Select a node $x$ from which to start the investigation, and, for each network $G_T$, execute the Algorithm~\ref{algo:Flow_Algo}.
    \item Store the results of the previous step in a table. Each row of the table represents a path in time period $T$. For each path at period $T$, we have the list of the weights of the edges involved in that path. If available, additional information of the nodes involved in the path can be included;
    \item The AFC analyst can now examine the table by filtering and sorting the paths. It is possible to filter the table by selecting all paths that end at a node $y$ and obtain the transaction flow from $x$ to $y$. If other information about the nodes is available, it can be used to further filter the results. For each time period $T$ and transaction flow, we can compute several time series metrics, such as the moving average or the exponentially weighted moving average. These metrics can be used for sorting the transaction flows, as described in Section \ref{subsec:timeseriesflows}. 
\end{enumerate}

In summary, by executing Algorithm~\ref{algo:Flow_Algo} on a properly constructed network, the AFC analyst will discover all the money paths of length \textit{n} between any two nodes and will be able to analyze statistically as time series, in order to identify potentially abnormal connections.

\section{Results}
\label{sec:results}

We test the potential of the \textit{WeirdFlows} tool by applying it to a dataset of 80 million cross-border transactions over 15 months provided by Intesa Sanpaolo (ISP), the largest Italian bank operating across Europe. The dataset\footnote{The data supporting the findings of this study is available from ISP upon request to AFC Digital Hub. Please note that restrictions for data availability apply. Researchers interested in having access to data for academic purposes will be asked to sign a non-disclosure agreement. Contact adh@pec.afcdigitalhub.com for further information.}, that includes all cross-border transactions involving ISP customers, either as sender or receiver, is described into details in~\cite{vilella2023anomaly}. The data can be aggregated either by IBANs, BIC codes or Countries - modeling transactions respectively between IBANs, BICs or nations at a progressively more coarse-grained resolution.

\subsection{Use Case: Money flows between prominent European Countries during the Ukrainian war}

In this section, we show an example of how the \textit{WeirdFlows} tool can be used in a financial crime investigation. The example involve networks with different types of node aggregation and show how \textit{WeirdFlows} can assist in the identification of complex patterns that can conceal financial fraud.
For privacy and security reasons, measures are taken in accordance with the legal provisions and the requirements of AFC Digital Hub consortium. In particular, transaction amounts are normalized to the maximum amount of the time series and BIC codes, that were already provided to the researchers in complete anonymisation, are further anonymised. Furthermore, all Country names will be anonymized as $C1, C2\ldots$ for the sake of data protection as requested by the data provider. 
\begin{figure}[ht!]
\centering
    \includegraphics[width=.83\linewidth]{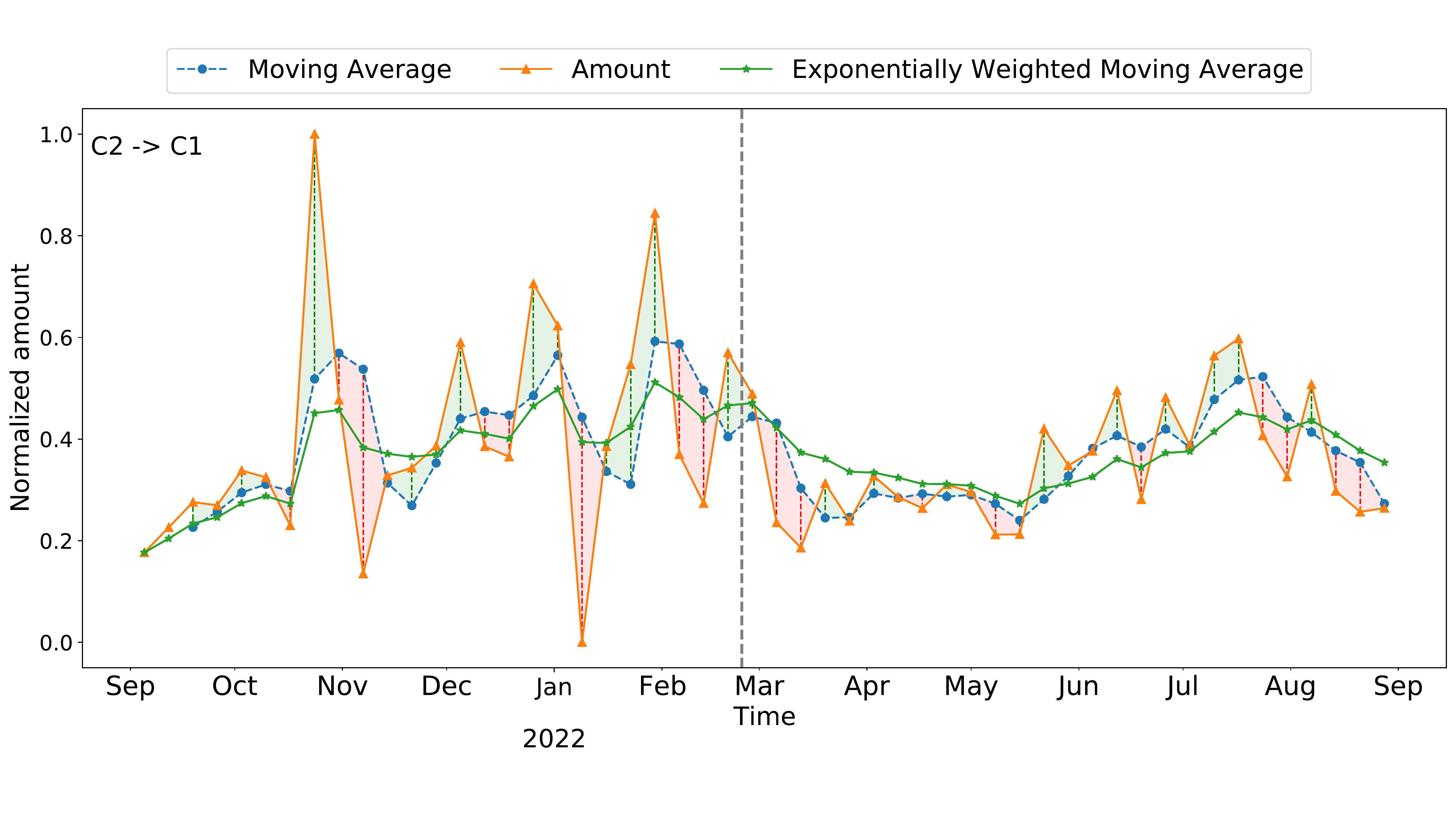}
  \caption{Direct transactions from C2 BICs to C1 BICs. The vertical grey dotted line represents the begin of the war in Ukraine (24 February 2022).}
  \label{fig:fig2}
\end{figure}

This use case investigates transaction flows between two prominent European countries at a macro scale, specifically focusing on the period between the outbreak of war in C3 and the subsequent imposition of economic sanctions by the EU.
Figure~\ref{fig:fig2} illustrates the weekly aggregated direct amounts transferred from C2 BICs to C1 BICs. The grey dashed line marks the start of the war in C3. The trend remains stable throughout the period, indicating that economic sanctions did not significantly alter the transaction amounts from C2 to C1. Figure~\ref{fig:fig3} displays the time series of $w(Flow^3(C2, C1))$, representing the maximum potential amount sent from C2 to C1 BICs through one intermediary. Both figures, especially before the war outbreak, show peaks but no discernible patterns.
To further investigate the issue, an AFC analyst can examine the intermediaries in the $Flow^3(C2, C1)$ transaction flow. There are 86 paths of maximum length 3 between C2 and C1. Table~\ref{tab:inter} highlights the intermediaries that exhibit the largest percentage increase in actual value over the expected moving average after February 24. As we can see, country C13 stands out with a 66\% increase: one can investigate this evidence by analyzing all the transactions that pass through C13. Having exogenous information helps: by deploying the pipeline of anomaly detection on financial graphs described in~\cite{vilella2023anomaly}, the BIC identified as BIC03C2 - belonging to Country C2 - was correctly identified as involved in malicious operations. If we use it as seed for WeirdFlows, analyzing all transactions between BIC03C2 and Country C1 that pass through country C13, we clearly see that the weight of the transaction flow shows an increasing trend after the outbreak of the war in Ukraine, peaking in the months of May and June.

\begin{minipage}{\textwidth}
  \begin{minipage}[b]{0.60\textwidth}
    \centering
    \includegraphics[width=1\linewidth]{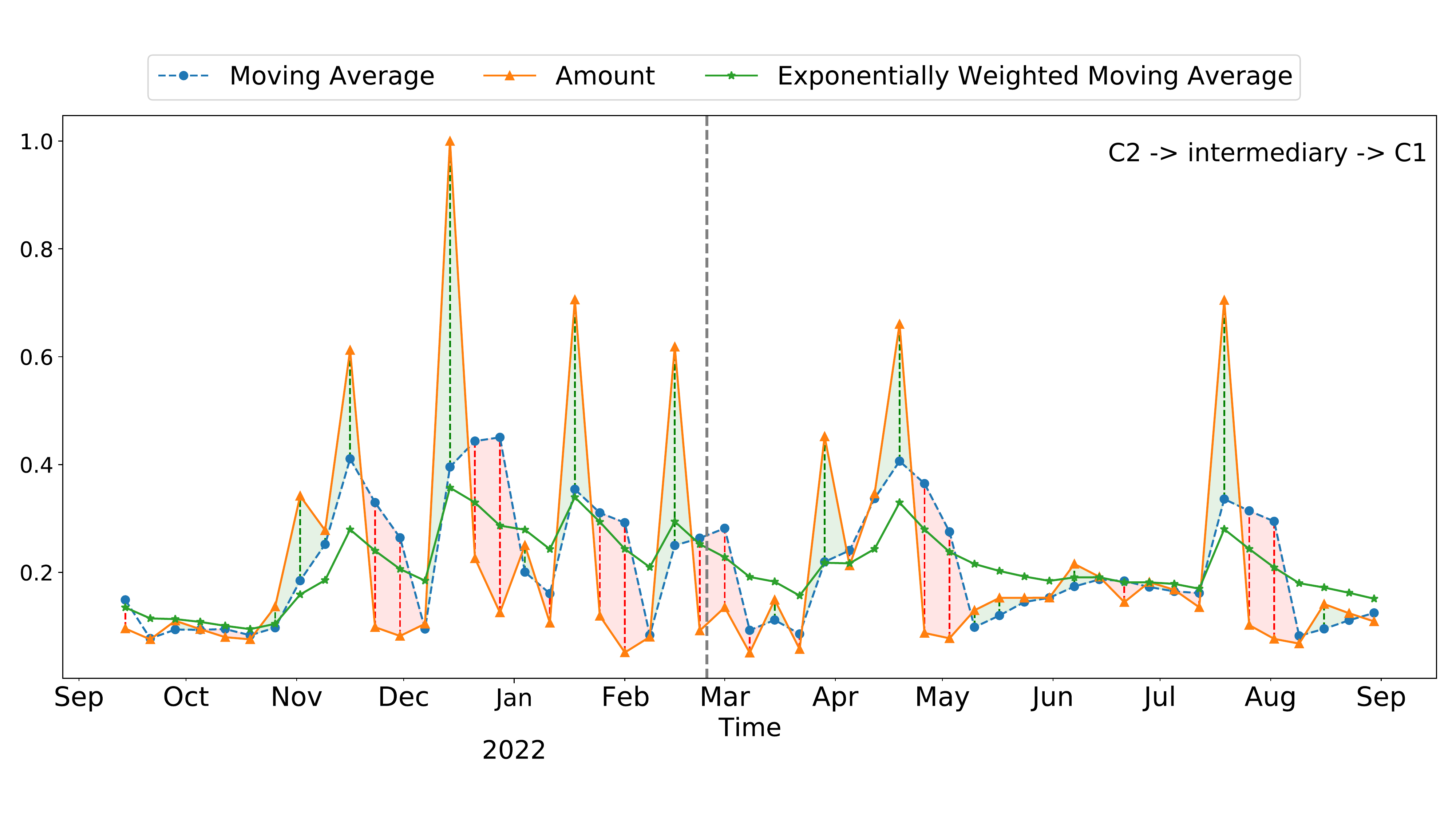}
    \captionof{figure}{Weight of the transaction flow $Flow^3(C2, C1)$, i.e., the hypothetical maximum amount of money sent from C2 to C1 through several payment lines, each with a maximum of one intermediary.}
    \label{fig:fig3}
  \end{minipage}
  \hfill
  \begin{minipage}[b]{0.38\textwidth}
    \centering
    \begin{tabular}{lr}
    \toprule
    Country &  Difference \\
    \midrule
         C4 &       0.427 \\
         C6 &       0.436 \\
         C5 &       0.493 \\
         C7 &       0.535 \\
         C8 &       0.539 \\
         C9 &       0.549 \\
         C10 &       0.604 \\
         C11 &       0.636 \\
         C12 &       0.662 \\
         C13 &       0.665 \\
    \bottomrule
    \end{tabular}
      \captionof{table}{Intermediaries in $Flow^3(C2, C1)$ that show the largest percentage increase between the expected value given by the moving average and the actual value.}
      \label{tab:inter}
    \end{minipage}
\end{minipage}

\begin{figure}[ht!]
\centering
    \includegraphics[width=.85\linewidth]{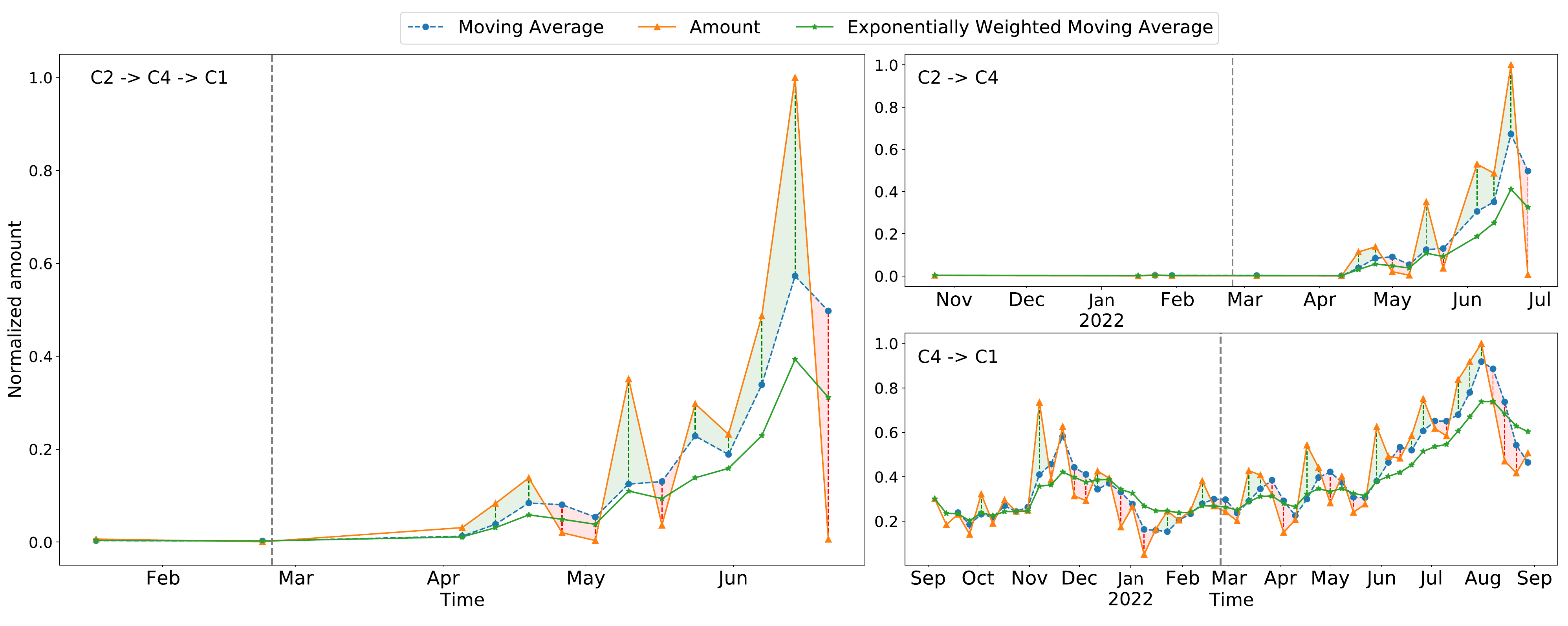}
  \caption{Transaction flow from C2 to C1 through C4.}
  \label{fig:fig4}
\end{figure}

\begin{figure}[ht!]
\centering
    \includegraphics[width=.88\linewidth]{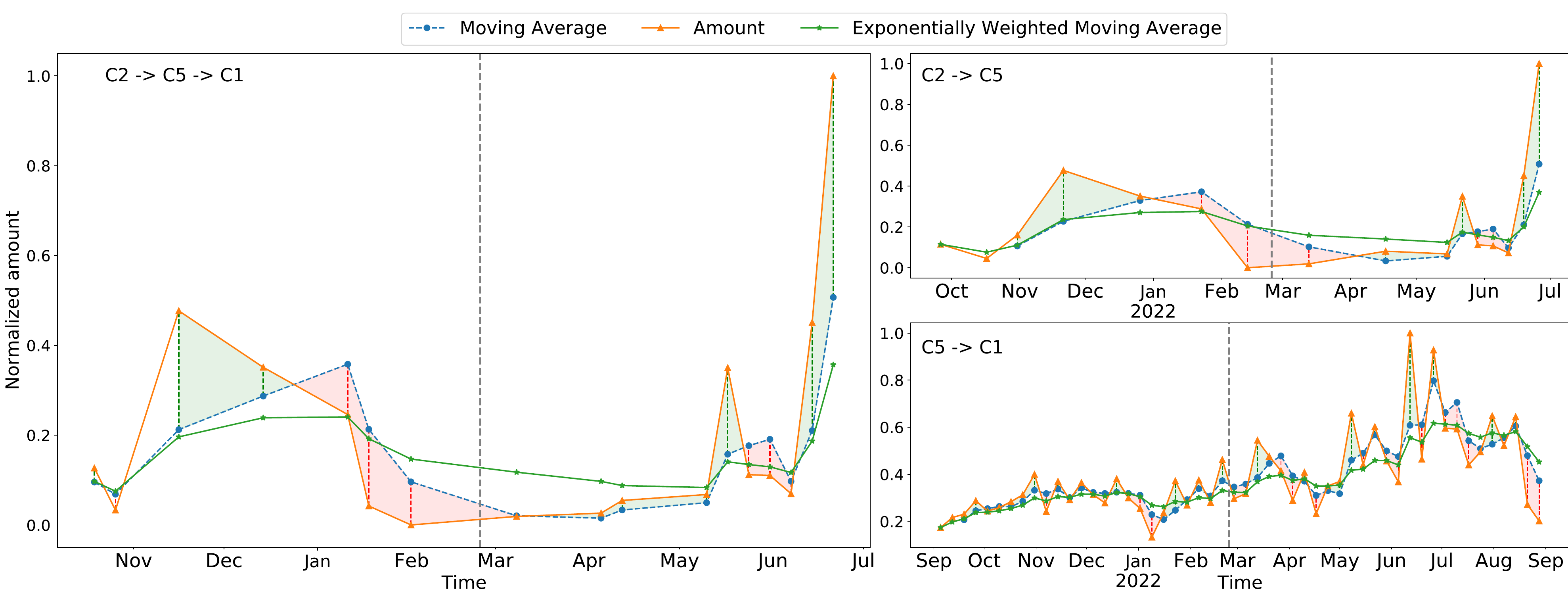}
  \caption{Transaction flow from C2 to C1 through C5.}
  \label{fig:fig5}
\end{figure}

Figures~\ref{fig:fig4} and~\ref{fig:fig5} depict transaction flows from C2 to C1 through C4 and C5, respectively. 

In Figures~\ref{fig:fig4} and~\ref{fig:fig5}, the left side shows the flow weights computed on weekly aggregated networks. On the right side, individual edges involved in the flows are depicted: C2 to C4 and C4 to C1 in Figure~\ref{fig:fig4} and C2 to C5 and C5 to C1 in Figure~\ref{fig:fig5}. From May 2022, the weight of $w(Flow^3(C2, C1))$ shows rapid and abnormal growth. Analyzing these individual edges, discovered through WeirdFlows, helps the analyst understand this growth. 
Incidentally, very similar patterns of intermediaries used to bypass international sanctions have been well documented\footnote{\url{https://www.ft.com/content/0fc846f7-aac8-4a34-a7dd-3b0615bce983}}.

\begin{figure}[ht!]
\centering
    \includegraphics[width=.75\linewidth]{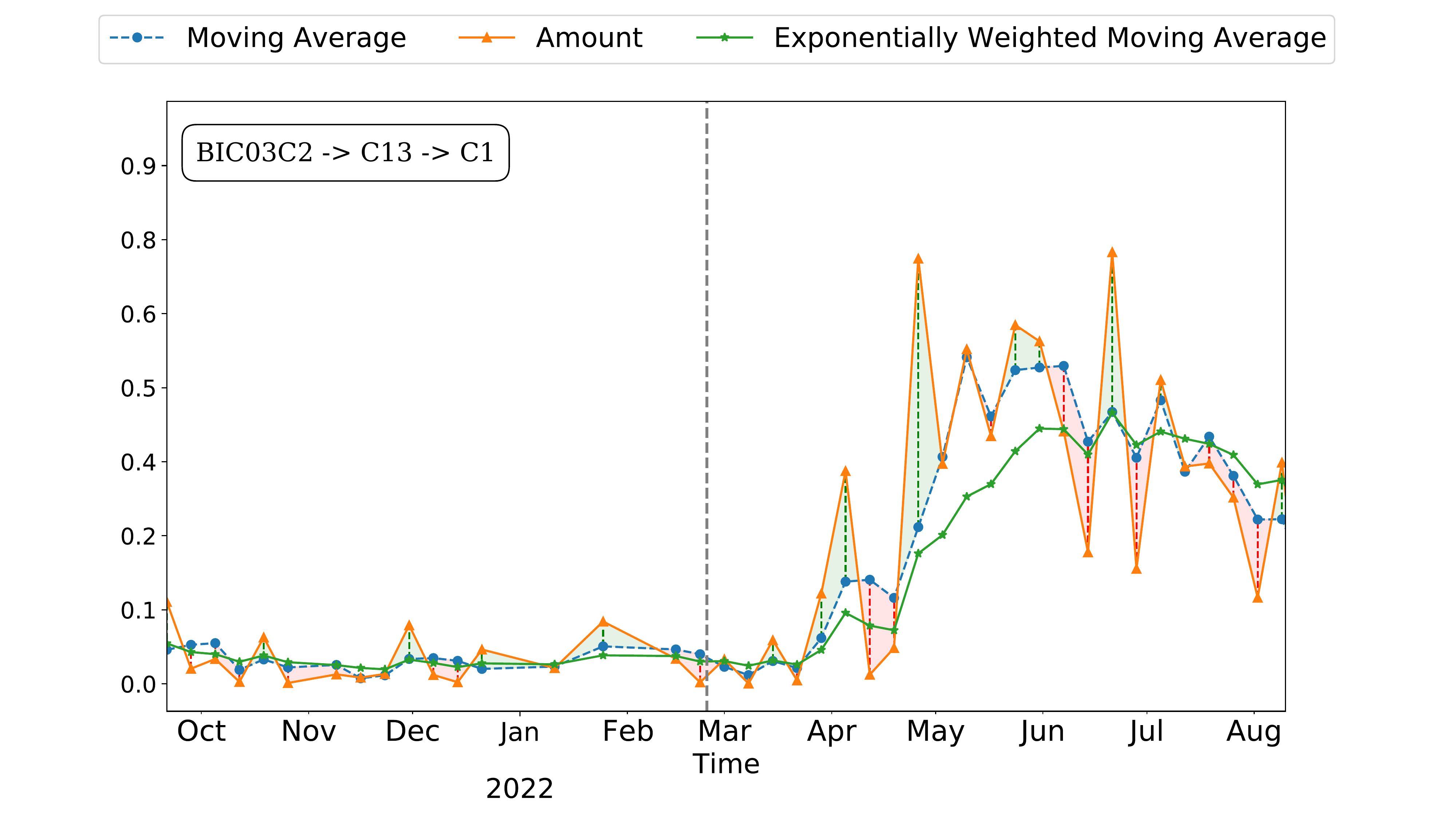}
  \caption{Transactions towards country C1, passing through intermediaries in country C13, originating from a specific BIC (BIC03C2) involved in malicious operations.}
  \label{fig:fig6}
\end{figure}

\bibliographystyle{splncs04}  
\bibliography{references}  

\end{document}